\documentclass[journal]{IEEEtran}
 \usepackage{cite}
\usepackage{multicol,lipsum}
\usepackage{amsmath,amssymb,amsfonts}
\usepackage{graphicx}
\usepackage{textcomp}
\usepackage{xcolor}
\usepackage{booktabs}
 \usepackage{hyperref}
\usepackage{enumerate}
\usepackage{algorithm}
\usepackage{multirow}
\usepackage{standalone}
\usepackage{tikz}
\usetikzlibrary{positioning}
\usetikzlibrary{decorations.markings}
\usepackage{float}
\makeatletter

\newcommand{\Rmnum}[1]{\expandafter\@slowromancap\romannumeral #1@}
\makeatother

\usepackage{graphicx}
  \usepackage{enumerate}
 \usepackage{caption}
\usepackage{subcaption}
 \usepackage{setspace}
\usepackage{verbatim}
   \usepackage{tabularx}
 \usepackage{makecell}
 \usepackage{hyperref}

\usepackage{algpseudocode}
\usepackage{algcompatible}
\ifCLASSINFOpdf
\else
\fi
 
\begin{document}
\title{A Progressive Codebook Optimization Scheme for Sparse Code Multiple Access in Downlink Channels}

\author{Tuofeng Lei, Qu Luo,~\IEEEmembership{Graduate Student Member,~IEEE,} Shuyan Ni,
        Shimiao Chen, Xin Song,
        and Pei Xiao,~\IEEEmembership{Senior Member,~IEEE}
        % <-this % stops a space
 
\thanks{Tuofeng Lei, Shuyan Ni, Shimiao Chen is with the Space Engineering University, Beijing 101416, China. (e-mail: Tuofenglei@163.com; daninini@163.com; 14291002@bjtu.edu.cn).}% <-this % stops a space
\thanks{Pei Xiao is with the State Key Laboratory of ISN, Xidian University,  Xi'an, 710071, China. email: pei.xiao@xidian.edu.cn.}% <-this % stops a space
\thanks{Qu Luo is with the National Key Laboratory of Science and  Technology on Communications, University of Electronic Science and Technology of China,  Chengdu  610054, China. email: ethan.luoqu@outlook.com.}% <-this % stops a space
\thanks{Xin Song is with the  Academy of Military Sciences, Beijing 100071, China. (e-mail: singlersx@163.com).}}
 
\maketitle
\begin{abstract}
 
Sparse code multiple access (SCMA) is a promising technique for enabling massive connectivity and high spectrum efficiency in future machine-type communication networks. However, its performance crucially depends on well-designed multi-dimensional codebooks. In this paper, we propose a novel progressive codebook optimization scheme that can achieve near-optimal performance over downlink fading channels. By examining the pair-wise error probability (PEP), we first derive the symbol error rate (SER) performance of the sparse codebook in downlink channels, which is considered as the design criterion for  codebook optimization. Then, the benchmark constellation group at a single resource element   is optimized with a sequential quadratic programming  approach. Next, we propose a constellation group reconstruction process to assign the sub-constellations in each resource element (RE) progressively. For the current RE, the assignment of the sub-constellations is designed by minimizing the error performance of the product distance of the superimposed codewords in previous REs. The design process involves both permutation and labeling of the sub-constellations in the benchmark constellation group. Simulation results show that the proposed codebooks exhibit significant performance gains over state-of-the-art codebooks in the low signal-to-noise ratio (SNR) region over various downlink fading channels.

\end{abstract}


% \fi
\IEEEpeerreviewmaketitle

\section{Introduction}
 \IEEEPARstart{T}{he} ever-growing demand for higher data rates, enhanced spectral efficiency, and massive connectivity  has driven the rapid evolution of wireless communication systems \cite{Liungma,ElbayoumiNOMA}.   To meet these stringent requirements, non-orthogonal multiple access (NOMA) has emerged as a promising technology for the future of wireless communication networks \cite{LuoError}. In contrast to traditional orthogonal multiple access (OMA), NOMA enables multiple users to effectively utilize the same resources in a non-orthogonal manner \cite{On_the_Design_of_Near_Optimal_Sparse_Code_Multiple_Access_Codebooks, Multidimensional_Constellations_for_Uplink_SCMA_Systems_A_Comparative_Study}.  Existing
NOMA methods are mainly categorized into power-domain 
\cite{Liungma} and code-domain NOMA \cite{LuoError}.  This paper focuses on a representative code domain NOMA (CD-NOMA) scheme called sparse code multiple access (SCMA), which has attracted significant research attention due to its excellent performance and low receiver complexity \cite{LuoE2E}.  \color{black}In SCMA, the incoming   message bits of each   user  are directly mapped to a multi-dimensional codeword chosen from a pre-defined codebook.  The codebook are intentionally designed with certain sparsity to effectively reduce the  decoding complexity of the  message passing algorithm (MPA). In addition, constellation shaping of multi-dimensional constellations  leads to a significant improvement in spectral efficiency when compared to other code-domain NOMA schemes \cite{On_Near_Optimal_Codebook_and_Receiver_Designs_for_MIMO_SCMA_Schemes}, such as low-density spreading code division multiple access (LDS-CDMA) and low-density signature-orthogonal frequency division multiplexing (LDS-OFDM).

\subsection{Related works}

SCMA codebook design has been studied extensively in
the past few years \cite{A_Partial_Gaussian_Tree_Approximation_(PGTA)_Detector_for_Random_Multiple_Access_Oriented_SCMA_Uplink_With_Codebook_Collisions,On_Near_Optimal_Codebook_and_Receiver_Designs_for_MIMO_SCMA_Schemes,SCMA_System_Design_With_Index_Modulation_via_Codebook_Assignment,Multidimensional_Constellation_Design_for_Spatial_Modulated_SCMA_Systems,Bit_Interleaved_Coded_SCMA_With_Iterative_Multiuser_Detection_Multidimensional_Constellations_Design,A_Low_Complexity_Codebook_Optimization_Scheme_for_Sparse_Code_Multiple_Access,Sparse_or_Dense_A_Comparative_Study_of_Code-Domain_NOMA_Systems,SCMA_Codebook_for_Uplink_Rician_Fading_Channels,SCMA_Codebook_Design,SCMA_for_downlink_multiple_access_of_5G_wireless_networks}. The overall design goal is to find a class of codebooks for low error rate or large spectral efficiency under  different channel conditions. Maximizing the minimum Euclidean distance of the superimposed codewords (MED-SC) and the minimum product distance of the superimposed codeword (MPD-SC) have been considered as the   key performance indicators in the additive white Gaussian noise (AWGN)    channels and downlink Rayleigh channels, respectively. To date, the AWGN codebook reported by Huang \cite{Downlink_SCMA_Codebook_Design_With_Low_Error_Rate_by_Maximizing_Minimum_Euclidean_Distance_of_Superimposed_Codewords} achieves the largest MED-SC, and the downlink Rayleigh codebook proposed by Chen has better performance in comparison with existing codebooks \cite{On_the_Design_of_Near_Optimal_Sparse_Code_Multiple_Access_Codebooks},\cite{A_Comprehensive_Technique_to_Design_SCMA_Codebooks}. In \cite{A_Low_Complexity_Codebook_Design_Scheme_for_SCMA_Systems_Over_an_AWGN_Channel}, MCs with low-projection number were proposed for downlink AWGN channels by maximizing the MED-SC.

Existing codebook design  typically follows a multi-stage approach   by first constructing a   mother constellation (MC), followed by certain  user-specific  operators, such as phase rotation and permutation 
\cite{SCMA_Codebook_Design,SCMA_for_downlink_multiple_access_of_5G_wireless_networks},
which are applied to the MC to obtain codebooks for multiple users.  The basic rationale behind this approach is to improve the MED-SC and MPD-SC, thus achieving lower error rate performance.
Based on this approach, various  MCs  have been reported   for their flexibility and simplicity 
\cite{SCMA_Codebook_for_Uplink_Rician_Fading_Channels,SCMA_Codebook_Design,SCMA_for_downlink_multiple_access_of_5G_wireless_networks,High_Dimensional_Codebook_Design_for_the_SCMA_DownLink,A_Suboptimal_Algorithm_for_SCMA_Codebook_Design_over_Uplink_Rayleigh_Fading_Channels,Designing_optimum_mother_constellation_and_codebooks_for_SCMA,A_Novel_Scheme_for_the_Construction_of_the_SCMA_Codebook}. 
\color{black}For example, in \cite{Designing_optimum_mother_constellation_and_codebooks_for_SCMA},
optimum MCs  were designed by maximizing the  MED-SC.  In \cite{A_Novel_Scheme_for_the_Construction_of_the_SCMA_Codebook}, the 3D
MC  was investigated to improve the error
performance by increasing available dimensional information.  In \cite{SCMA_Codebook_for_Uplink_Rician_Fading_Channels}, the design
criteria of the SCMA codebook for the uplink Rician fading channels was investigated, and a mother constellation based
codebook was used to achieve better performance.  Moreover,  the author in \cite{LuoLPSCMA} studied the codebook design in Ricain fading channels, where a novel class of low-projection SCMA codebooks for ultra-low decoding complexity  was  developed. 
 In 
\cite{SCMA_Codebook_Design_Based_on_Uniquely_Decomposable_Constellation_Groups}, an
uniquely decomposable constellation group based codebook design approach was proposed to improve the error performance. In \cite{Design_of_SCMA_Codebooks_using_Differential_Evolution}, differential evolution optimization was adopted for minimizing the symbol error rates (SER) of the SCMA system.   In \cite{On_the_Design_of_Near_Optimal_Sparse_Code_Multiple_Access_Codebooks}, near optimal SCMA codebooks were proposed in both AWGN   and Rayleigh fading channels. It should be noted that  the optimization of the best rotation angles and the calculation of the Euclidean distance of the superimposed codewords  in these works generally incur   high computation complexity.

\subsection{Motivations and Contributions}
As pointed out in \cite{On_the_Design_of_Near_Optimal_Sparse_Code_Multiple_Access_Codebooks}, the superimposed constellation at each RE also has a great impact on the error rate performance.  In general, the constellation at each RE  forms a constellation group. Since many existing codebooks choose different MCs and rotation angles \cite{High_Dimensional_Codebook_Design_for_the_SCMA_DownLink,A_Suboptimal_Algorithm_for_SCMA_Codebook_Design_over_Uplink_Rayleigh_Fading_Channels,Designing_optimum_mother_constellation_and_codebooks_for_SCMA,A_Novel_Scheme_for_the_Construction_of_the_SCMA_Codebook,SCMA_Codebook_Design_Based_on_Uniquely_Decomposable_Constellation_Groups,Design_of_SCMA_Codebooks_using_Differential_Evolution,A_Low_Complexity_Codebook_Design_Scheme_for_SCMA_Systems_Over_an_AWGN_Channel,Design_of_SCMA_Codebooks_Based_on_Golden_Angle_Modulation,Design_and_Analysis_of_SCMA_Codebook_Based_on_Star_QAM_Signaling_Constellations,Design_of_Power_Imbalanced_SCMA_Codebook}, the resultant constellation groups at  each RE may also be different. For example, the constellation groups can be formed by the  Golden angle modulation (GAM) constellation and star quadrature amplitude modulation (Star-QAM)  constellation in \cite{Design_of_SCMA_Codebooks_Based_on_Golden_Angle_Modulation} and \cite{Design_and_Analysis_of_SCMA_Codebook_Based_on_Star_QAM_Signaling_Constellations}, respectively. However, the multi-stage approach is generally considered a sub-optimal approach, as the superimposed constellation depends crucially on the choice of MC and rotation angles.  An alternative and more ambitious approach is to directly design the superimposed constellation at each RE with desirable characteristics instead of following a multi-stage  based approaches.

Driven by this rationale, we propose to directly design the superimposed constellation at a single RE, where the multiple constellations that lead to the superimposed constellation are referred to as the benchmark constellation group. Subsequently, we suggest a reconstruction process by progressively assigning the benchmark constellation group to each RE. At each step, the assignment of the benchmark constellation group for the current RE is determined by minimizing the error performance of the product distance among superimposed codewords from previous REs. Moreover, unlike many existing SCMA papers that focus on codebook design for AWGN and Rayleigh fading channels, we design SCMA codebooks for Rician and Nakagami-$m$ fading channels, which are prevalent  in practical networks. The major contributions of the paper are summarized as follows:
\color{black}
\begin{itemize}
    \item We derive the SER performance of  SCMA systems over different fading channels by analyzing the pair-wise error probability (PEP).    The related properties and codebook design criteria in each fading channel are also investigated and analyzed.
     \item We propose a novel progressive codebook optimization scheme based on a benchmark constellation group. Specifically, we optimize the benchmark constellation group at a single RE using a sequential quadratic programming (SQP) approach. Next, we introduce a constellation group reconstruction process to progressively assign the sub-constellations of the benchmark constellation group in each RE.

     \item  We conduct extensive simulation results to demonstrate the superiority of the proposed codebooks over different fading channels. The proposed codebook exhibits  the notable improvements in terms of error performance under lower SNRs over the  existing  near optimal codebooks. 
   
\end{itemize}

\subsection{Organization}
The remainder of this paper is organized as follows: Section \Rmnum{2} introduces the model of the downlink SCMA system. Section \Rmnum{3} provides the SER formula of the sparse multidimensional constellation under different fading channels. The detailed construction procedures of the codebooks are presented in Section \Rmnum{4}. Then, in Section \Rmnum{5}, simulation results are provided to show the SER performance of the proposed codebooks. Finally, Section \Rmnum{6} concludes the paper.

 \subsection{Notations}

  $\mathbb{C}^{k\times n}$ and $\mathbb{B}^{k\times n}$ denote the $(k\times n)$-dimensional complex and binary matrix spaces, respectively.  Scalars, vectors and matrices are distinguished by normal, lowercase bold and uppercase bold fonts.  
  ${\mathrm{|}}\cdot {\mathrm{|}}$ and ${\mathrm{||}}\cdot {\mathrm{||}}$ denote the   absolute value, and the $\mathcal   \ell _{2}-\mathrm {norm}$, respectively.  
$ \mathcal{CN}\left (0, 1\right )$  denotes complex Gaussian distribution with  zero-mean and unit-variance. 

\section{ SCMA Communication Model}

\begin{figure}[t]
    \centering
    \includegraphics[width=1.0\linewidth]{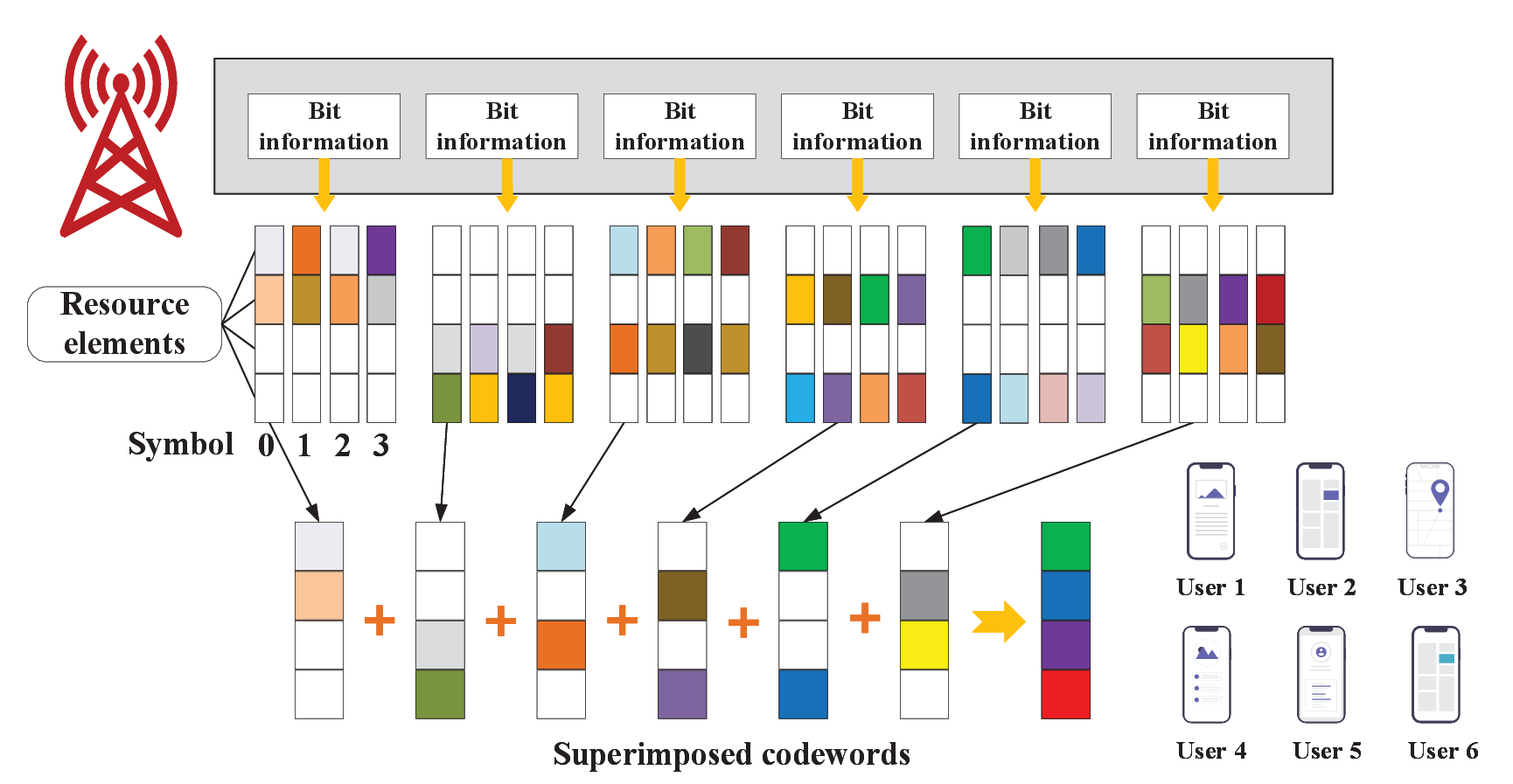}
    \caption{{An example of SCMA mapping process.}}
    \label{fig:The Example of SCMA Encoding}
\end{figure}

We consider a  $K \times L$  SCMA system, where $L $ users communicate over $K$ REs for multiple access.  The overloading factor of SCMA system is defined by $\xi = L/K > 100 \%$, which  indicates that the number of users that
concurrently communicate is larger than the total number of orthogonal resources. Each user is assigned with  a unique codebook, denoted by  $  {\mathcal {X}} _{l}=\left\{\mathbf{x}_{l, 1}, \mathbf{x}_{l, 2}, \ldots, \mathbf{x}_{l, M}\right\}   \in \mathbb {C}^{K \times M}, l \in\{1,2, \ldots, L\}$, consisting of $M$ codewords with a dimension of $K$.  During transmissions,  each user   maps $\log_2\left(M\right)$   binary bits to a length-$K$ codeword $ \mathbf {x} _{l}$ drawn from the $  {\mathcal {X}}_{l}$. 
The SCMA mapping process can be written as
\begin{equation}
    \begin{matrix}
   {{f}_{l}}:{{\mathsf{\mathbb{B}}}^{lo{{g}_{2}}M\times 1}}\to {{\mathcal{X}}_{l}}, & \text{i}\text{.e}.,{{\mathbf{x}}_{l}}={{f}_{l}}  \\
    \end{matrix}\left( {{\mathbf{b}}_{l}} \right),\
\end{equation}
where ${{\mathbf{b}}_{l}}\in {{\mathsf{\mathbb{B}}}^{lo{{g}_{2}}M\times 1}}$ denotes the bits data of the $l\text{th}$ user, $M$ represents the codebook size, and ${{\mathbf{x}}_{l}}={{\left[ {{x}_{1,l}},{{x}_{2,l}},\cdots ,{{x}_{K,l}} \right]}^{T}}$ denotes the transmitted codewords in the $l\text{th}$ user’s codebook ${ {\mathcal{X}}_{l}}$, and ${{x}_{k,l}}$ is the transmitted codeword of the $l\text{th}$ user at the $k\text{th}$ RE.   Fig. \ref{fig:The Example of SCMA Encoding} shows an example of the SCMA mapping process.

\begin{figure}[!t]
    \centering
    \includegraphics[width=0.8\linewidth]{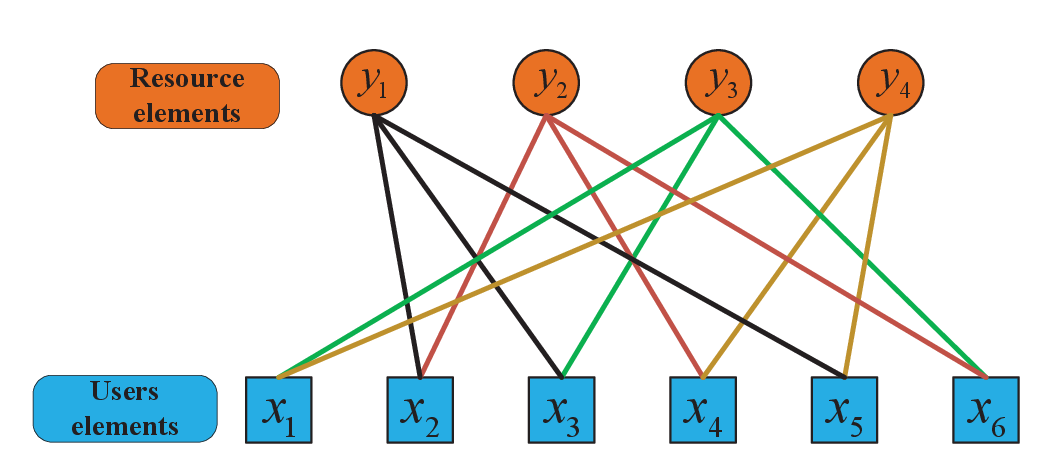}
    \caption{{The factor graph representation of the indicator matrix.}}
    \label{fig:The Tanner representation of the factor graph matrix}
\end{figure}

The $K$-dimensional codewords are sparse vectors with $d_v$ non-zero elements. The sparse structure of the $L$ codebooks can be represented by an   indicator matrix   $\mathbf {F}  \in \mathbb {B}^{K\times L}$. An element of  ${\bf{F}}$ is defined as ${f_{k,l}}$ which takes the value of $1$ if and only if  the $k$th RE is occupied by  the $l$ user, and $0$ otherwise. The   indicator matrix is constructed by the progressive edge-growth (PEG) algorithm to attain large girths \cite{A_Novel_Scheme_for_the_Construction_of_the_SCMA_Codebook}. For  the SCMA system with $K=4$, $L=6$ and $d_v=2$, the indicator matrix is given as
\begin{equation}
    \textbf{F} =\left[ \begin{matrix}
   0 & 1 & 1 & 0 & 1 & 0  \\
   0 & 1 & 0 & 1 & 0 & 1  \\
   1 & 0 & 1 & 0 & 0 & 1  \\
   1 & 0 & 0 & 1 & 1 & 0  \\
    \end{matrix} \right],
\end{equation}
where the corresponding factor graph representation is  shown in Fig. \ref{fig:The Tanner representation of the factor graph matrix}. We further define the set $\phi _k=\lbrace l:  {f}_{k,l} = 1 \rbrace$ consisting of all the  users colliding over  RE $k$ and the   number of   users collides over  a RE is given as $d_f = \vert\phi _k \vert $.

\color{black} Denote ${{h}_{k}}$ by the channel coefficient at the $k\text{th}$ RE, the received signal at the $k\text{th}$ RE can be written as
\begin{equation}
    \begin{matrix}
   {{y}_{k}}={{h}_{k}}\sum\limits_{l=1}^{L}{{{x}_{k,l}}+{{z}_{k}},} & {{x}_{k,l}}\in { {\mathcal{X}}_{l}}  \\
    \end{matrix},
\end{equation}

\noindent
where ${{z}_{k}}$ denotes the additive white Gaussian noise at the $k\text{th}$  RE that obeys complex circularly symmetric Gaussian random variables $\mathcal{C}\mathcal{N}\left( 0,{{N}_{0}} \right)$.

\section{Error Performance of the Sparse Codebook in Different Fading Channels}

In downlink SCMA systems, users' data are first superimposed at the base station side, which constitutes the superimposed constellation ${{\mathcal{X}}_{\text{sup}}}$ with size of $K  \times M^L$.
Denote $\mathbf c = \sum_{l=1}^{L} \mathbf x_l \in {{\mathcal{X}}_{\text{sup}}}$ as the transmitted superimposed codeword of $L$ users.   Assume the codeword is erroneously decoded as  $\mathbf c_j$  when  $\mathbf c_i$  is transmitted, where   $\mathbf c_i \neq \mathbf c_j$.  Then,   the conditional  PEP   between be two distinct codewords $\mathbf c_i $ and $\mathbf c_j$    can be calculated by
\begin{equation}
    \Pr \left( \left. {{\mathbf{c}}_{i}}\to {{\mathbf{c}}_{j}} \right|\mathbf{h} \right)=Q\left( \sqrt{\frac{{{E}_{s}}}{2{{N}_{0}}}d_{\text{sup}}^{2}\left( {{\mathbf{c}}_{i}},{{\mathbf{c}}_{j}} \right)} \right),
\end{equation}
where $Q\left( x \right)={1}/{\sqrt{2\pi }}\;\int_{x}^{\infty }{\exp \left( -{{{t}^{2}}}/{2}\; \right)dt}$ denotes the $Q$-function \cite{Performance_and_design_of_space_time_coding_in_fading_channels}, $\mathbf{h}=[ 
   {{h}_{1}},{{h}_{2}}   \ldots,   {{h}_{K}} ]  $ denotes the channel coefficient vector,  and  $d_{\text{sup}}^{2}\left( {{\mathbf{c}}_{i}},{{\mathbf{c}}_{j}} \right)$ denotes the Euclidean distance between the superimposed codewords which takes the following expression
\begin{equation}
    d_{\text{sup}}^{2}\left( {{\mathbf{c}}_{i}},{{\mathbf{c}}_{j}} \right)=\sum\limits_{k=1}^{K}{{{\left| {{h}_{k,l}}\left( {{c}_{k,i}}-{{c}_{k,j}} \right) \right|}^{2}}},
\end{equation}

\noindent
where ${{c}_{k,i}}=\sum\nolimits_{l=1}^{L}{{{x}_{k,l}}}$ is the $i\text{th}$ superimposed codeword  at the $k\text{th}$ RE, and ${h}_{k,l}$ represents the channel coefficients of the $l\text{th}$ user at the $k\text{th}$ RE.  Then, following the Chernoff bound  \cite{Performance_and_design_of_space_time_coding_in_fading_channels}, the conditional PEP can be expressed as
\begin{equation}
    \Pr \left( \left. {{\mathbf{c}}_{i}}\to {{\mathbf{c}}_{j}} \right|\mathbf{h} \right)\le \frac{1}{2}\exp \left( -\frac{{{E}_{s}}}{4{{N}_{0}}}d_{\text{sup}}^{2}\left( {{\mathbf{c}}_{i}},{{\mathbf{c}}_{j}} \right) \right).
\end{equation}
\color{black}

With the aid of the MGF, the PEP of the superimposed codewords can be easily obtained. Denote the instantaneous SNR as $\gamma ={{\left| {{h}_{k}} \right|}^{2}}\left( {{{E}_{s}}}/{{{N}_{0}}}\; \right)$, and the average SNR as $\bar{\gamma }=\mathbb{E}\left[ \gamma  \right]$, where $\bar{\gamma }$ is the mean square of the instantaneous SNR. Thus, the MGF associated with the instantaneous SNR is defined as \cite{Digital_communication_over_fading_channels}
\begin{equation}
    {{\mathcal{M}}_{\gamma }}\left( s \right)=\int_{0}^{\infty }{p\left( \gamma  \right)}{{e}^{s\gamma }}d\gamma ,
\end{equation}
\noindent
where $ p\left( \gamma  \right)$ is the PDF of the $\gamma $. The unconditional PEP can be obtained by   averaging over the channel distribution
\begin{equation}
\begin{aligned}
&\mathrm{Pr}\left(\mathbf{c}_{i}\right.  \to\mathbf{c}_j)  =  \mathbb E_{\mathbf h} \left \{\Pr \left( \left. {{\mathbf{c}}_{i}}\to {{\mathbf{c}}_{j}} \right|\mathbf{h} \right) \right\}\\
&\leq\frac12\int_0^\infty\exp\left(-\frac{E_s}{4N_0}\sum_{k=1}^K\Bigl|h_{k,l}\bigl(c_{k,i}-c_{k,j}\bigr)\Bigr|^2\right)p(h) {d} h, \\
&\leq\frac12\prod_{k=1}^K\mathcal{M}_\gamma(-s_k),
\end{aligned}
\end{equation}
where ${{s}_{k}}={{{\left| {{c}_{k,i}}-{{c}_{k,j}} \right|}^{2}}}/{4}\;$.   The average SER can be written as \cite{LuoSSD}
\begin{equation}
\label{ASER}
    {{P}_{e}}\le \frac{1}{{{M}^{L}}}\sum\limits_{{\mathbf{c}_{i}}\in {{\mathcal{X}}_{\text{sup}}}}{\sum\limits_{{\mathbf{c}_{j}}\in {{\mathcal{X}}_{\text{sup}}}\backslash \left\{ {\mathbf{c}_{i}} \right\}}{\frac{1}{2}\prod\limits_{k=1}^{K}{{{\mathcal{M}}_{\gamma }}\left( -{{s}_{k}} \right)}.}}
\end{equation}

In the following sections, we derive the average  SER  of SCMA over different fading channels based on   (\ref{ASER}).

\subsection{Rayleigh Fading Channel}

The Rayleigh channel model is suitable for characterizing  the urban environment with dense buildings but no direct exposure. The distribution of Rayleigh fading is the ${{\ell }_{2}}$ norm of two zero-mean independent Gaussian random variables, and the PDF of the Rayleigh distribution can be written as
\begin{equation}
    p\left( z \right)=\begin{matrix}
   \frac{z}{\sigma _{Ra}^{2}}\exp \left( -\frac{{{z}^{2}}}{2\sigma _{Ra}^{2}} \right), & z>0  \\
\end{matrix},
\end{equation}

\noindent
where $\sigma _{Ra}^{2}$ is the scatter component of the Rayleigh distribution, the mean square of the Rayleigh distribution is given by $\mathbb{E}\left[ {{z}^{2}} \right]=2\sigma _{Ra}^{2}$ \cite{Digital_Communications}, and the MGF of the instantaneous SNR $\gamma $ is \cite{Digital_communication_over_fading_channels}
\begin{equation}
    \begin{matrix}
   {{\mathcal{M}}_{\gamma }}\left( -s \right)=\frac{1}{1+s\bar{\gamma }}, & s>0  \\
    \end{matrix}.
\end{equation}

The PEP of the superimposed codeword in the Rayleigh fading channel can be calculated as
\begin{equation}
    \Pr \left( {{\mathbf{c}}_{i}}\to {{\mathbf{c}}_{j}} \right)\le \frac{1}{2}{{\prod\limits_{k=1}^{K}{\left( 1+\frac{2\sigma _{Ra}^{2}{{E}_{s}}}{4{{N}_{0}}}{{\left| {{c}_{k,i}}-{{c}_{k,j}} \right|}^{2}} \right)}}^{-1}}.
\end{equation}

At  high SNRs with a large scatter component $\sigma _{Ra}^{2}$, it can be observed that maximizing the product distance of the superimposed codewords ${{d}_{p}}=\prod\nolimits_{k=1}^{K}{{{\left| {{c}_{k,i}}-{{c}_{k,j}} \right|}^{2}}}$ helps improve error performance. In the low SNR region with a small scatter component $\sigma _{Ra}^{2}$, the Euclidean distance of the superimposed codeword ${{d}_{e}}=\sum\nolimits_{k=1}^{K}{{{\left| {{c}_{k,i}}-{{c}_{k,j}} \right|}^{2}}}$ also helps improve error performance. 

\subsection{Rician Fading Channel}
The Rician fading model is used to characterize  channels with direct line-of-sight (LoS) waves. The distribution of the Rician fading can be modeled as the ${{\ell }_{2}}$ norm of two non-zero mean independent Gaussian random variables and the PDF of the Rician distribution is given by \cite{LuoLPSCMA}
\begin{equation}
    p\left( z \right)=\begin{matrix}
   \frac{z}{\sigma _{Ri}^{2}}{{I}_{0}}\left( \frac{uz}{\sigma _{Ri}^{2}} \right)\exp \left( -\frac{{{z}^{2}}+{{u}^{2}}}{2\sigma _{Ri}^{2}} \right), & z>0  \\
    \end{matrix},
\end{equation}
\noindent
where $u$ and $\sigma _{Ri}^{2}$ are the LoS and the scatter components respectively, $\mathcal{K}={{{u}^{2}}}/{2\sigma _{Ri}^{2}}\;$ is defined as the Rician factor, and ${{I}_{0}}$ is the modified Bessel function of the first kind. The mean square of the Rician distribution can be calculated as $\mathbb{E}\left[ {{z}^{2}} \right]=2\sigma _{Ri}^{2}\left( 1+\mathcal{K} \right)$ \cite{Digital_Communications}, and the MGF of the instantaneous SNR $\gamma $ in the Rician distribution is given by \cite{Digital_communication_over_fading_channels}

\begin{equation}
    \begin{matrix}
   {{\mathcal{M}}_{\gamma }}\left( -s \right)=\frac{1+\mathcal{K}}{1+\mathcal{K}+s\bar{\gamma }}\exp \left( -\frac{\mathcal{K}s\bar{\gamma }}{1+\mathcal{K}+s\bar{\gamma }} \right), & s>0  \\
   \end{matrix}.
\end{equation}
Then, the PEP of the superimposed codeword in the Rician fading channel can be calculated
\begin{equation}
\begin{aligned}
\operatorname*{Pr}(\mathbf{c}_{i}\to\mathbf{c}_{j})& \leq\frac{1}{2}\prod_{k=1}^{K}\frac{1}{1+\frac{2\sigma_{Ri}^{2}E_{s}}{4N_{0}}\big|c_{k,i}-c_{k,j}\big|^{2}}  \\
&\times\exp\left(-\frac{\mathcal{K}\frac{2\sigma_{Ri}^{2}E_{s}}{4N_{0}}\big|c_{k,i}-c_{k,j}\big|^{2}}{1+\frac{2\sigma_{Ri}^{2}E_{s}}{4N_{0}}\big|c_{k,i}-c_{k,j}\big|^{2}}\right),
\end{aligned}
\end{equation}

\noindent
where we define the PEP  of the scatter component and the LoS component respectively as
\begin{equation}
    {{\Pr }_{\text{scatter}}}=\prod\limits_{k=1}^{K}{\frac{1}{1+\frac{2\sigma _{Ri}^{2}{{E}_{s}}}{4{{N}_{0}}}{{\left| {{c}_{k,i}}-{{c}_{k,j}} \right|}^{2}}}},
\end{equation}

\begin{equation}
    {{\Pr }_{\text{LoS}}}=\prod\limits_{k=1}^{K}{\exp \left( -\frac{\mathcal{K}\frac{2\sigma _{Ri}^{2}{{E}_{s}}}{4{{N}_{0}}}{{\left| {{c}_{k,i}}-{{c}_{k,j}} \right|}^{2}}}{1+\frac{2\sigma _{Ri}^{2}{{E}_{s}}}{4{{N}_{0}}}{{\left| {{c}_{k,i}}-{{c}_{k,j}} \right|}^{2}}} \right)}.
\end{equation}

Compared with the PEP in Rayleigh fading channel \cite{Design_of_Power_Imbalanced_SCMA_Codebook,A_Comprehensive_Technique_to_Design_SCMA_Codebooks}, SCMA achieves better error performance in the Rician fading channel from the LoS component ${{\Pr }_{\text{LoS}}}$. When the SNR and the scatter component $\sigma _{Ri}^{2}$ is sufficiently large, the PEP for the Rician fading channel can be simplified as
\begin{equation}
    \Pr \left( {{\mathbf{c}}_{i}}\to {{\mathbf{c}}_{j}} \right)\le \frac{1}{2}\prod\limits_{k=1}^{K}{{{\left( \frac{2\sigma _{Ri}^{2}{{E}_{s}}}{4{{N}_{0}}}{{\left| {{c}_{k,i}}-{{c}_{k,j}} \right|}^{2}} \right)}^{-1}}\exp \left( -\mathcal{K} \right)}.
\end{equation}
Meanwhile, it can be observed that maximizing the product distance of the superimposed codewords, $d_p$, is a crucial factor for the performance of the Rician fading channel. While in lower SNR with a larger Rician factor $\mathcal{K}$ (a small scatter component $\sigma _{Ri}^{2}$ or a larger LoS component $u$), observing from the LoS component ${{\Pr }_{\text{LoS}}}$ in (17), we can find that maximizing the Euclidean distance of the superimposed codeword ${{d}_{e}}$ also helps improve the performance.

\subsection{Nakagami-$m$ Fading Channel}
Nakagami-$m$ is a multipath channel model that can better control the degree of fading. Moreover, the Nakagami-$m$ distribution can be considered as the square root of the sum of squares of $2m$ independent zero mean Gaussian variates \cite{On_the_multivariate_Nakagami_m_distribution_with_exponential_correlation}. When the Nakagami-$m$ fading parameter $m$ is 1, the Nakagami-$m$ model is reduced to the Rayleigh model. The PDF of the Nakagami-$m$ fading model can be expressed as
\begin{equation}
    p\left( z \right)=\begin{matrix}
   \frac{2}{\Gamma \left( m \right)}{{\left( \frac{m}{\text{ }\!\!\Omega\!\!\text{ }} \right)}^{m}}{{z}^{2m-1}}\exp \left( -\frac{m{{z}^{2}}}{\text{ }\!\!\Omega\!\!\text{ }} \right), & z>0  \\
\end{matrix},
\end{equation}

\noindent
where $\text{ }\!\!\Omega\!\!\text{ }$ is the scale parameter, $m$ is the Nakagami-$m$ fading parameter, which has the range of $m\ge 0.5$, and $\Gamma \left( m \right)$ is the gamma function which can be written as
\begin{equation}
    \Gamma \left( m \right)=\int_{0}^{\infty }{{{t}^{m-1}}{{e}^{-t}}}dt.
\end{equation}
The mean square of the Nakagami-$m$ distribution can be calculated as $\mathbb{E}\left[ {{z}^{2}} \right]=\text{ }\!\!\Omega\!\!\text{ }\text{}$ \cite{Digital_Communications}, and the MGF of the instantaneous SNR $\gamma $ in the Nakagami-$m$ fading channel is given by \cite{Digital_communication_over_fading_channels}
\begin{equation}
    \begin{matrix}
   {{\mathcal{M}}_{\gamma }}\left( -s \right)={{\left( 1+\frac{s\bar{\gamma }}{m} \right)}^{-m}}, & s>0  \\
    \end{matrix}.
\end{equation}
The PEP of the superimposed codeword in the Nakagami-$m$ fading channel can be written as 
\begin{equation}
    \Pr \left( {{\mathbf{c}}_{i}}\to {{\mathbf{c}}_{j}} \right)\le \frac{1}{2}{{\prod\limits_{k=1}^{K}{\left( 1+\frac{\text{ }\!\!\Omega\!\!\text{ }\frac{{{E}_{s}}}{4{{N}_{0}}} {{\left| {{c}_{k,i}}-{{c}_{k,j}} \right|}^{2}}}{m} \right)}}^{-m}},
\end{equation}

\noindent
where one can see that maximizing the product distance of the superimposed codewords ${{d}_{p}}$ helps improve the error performance only when the ratio of the ${\text{ }\!\!\Omega\!\!\text{ }}/{m}\;$ and the SNR are large. While in the low SNR region or the ratio of the ${\text{ }\!\!\Omega\!\!\text{ }}/{m}\;$ is small, maximizing the Euclidean distance of the superimposed codeword ${{d}_{e}}$ can improve the error rate performance.

\section{The Proposed Progressive Codebook Optimization Scheme}

This section presents the detailed design of the proposed progressive codebook optimization scheme. First, the SER of the SCMA is adopted as the design criteria to obtain the near optimal codebook. Then, a progressive codebook optimization scheme is proposed  for different   channel models. 
\subsection{The Optimization of the Benchmark Constellation Group}

%\subsubsection{Constellation superposition} 
Denote $\mathcal{A}_0 \in \mathbb C^{M  \times 1}$ as the  benchmark constellation. In existing 
SCMA works, $\mathcal{A}_0$ is employed to design the multi-dimensional codebooks for different users.  Specifically, the  non-zero dimensions of the $l$th user's codebook can be generated by applying the user-specific operators, such as  phase rotation, to $\mathcal{A}_0 $.  Owning to the sparsity of the factor graph matrix,  the number of users superimposed on one RE,  denoted by ${{d}_{f}}$, is less than the number of users. Therefore,  ${{d}_{f}}$ distinct rotation angles  are sufficient to distinguish the superimposed codewords. 
 The codebooks for the $L$ users  can be represented by the signature matrix  $\mathbf{S}_{K\times L}$. For example, the   following signature matrix can be employed for efficient codebook construction 
  \begin{equation} 
 \label{signature_46}
 \small
 {{\mathbf{S}}_{4\times 6}}=\left[ \begin{matrix}
   0  & \mathcal{A}_0e^{j\theta_3} & \mathcal{A}_0e^{j \theta_1} & 0 &  \mathcal{A}_0e^{j\theta_2} & 0  \\
   \mathcal{A}_0e^{j\theta_2} & 0 & \mathcal{A}_0e^{j\theta_3} & 0 & 0 & \mathcal{A}_0e^{j\theta_1}\\
   0 & \mathcal{A}_0e^{j\theta_2}& 0 & \mathcal{A}_0e^{j\theta_1} & 0 & \mathcal{A}_0e^{j\theta_3} \\
   \mathcal{A}_0e^{j\theta_1} & 0 & 0 & \mathcal{A}_0e^{j\theta_2} & \mathcal{A}_0e^{j\theta_3} & 0  \\
\end{matrix} \right].
  \end{equation}

 Based on (\ref{signature_46}), the superimposed constellation on a RE  can be obtained  by 
   \begin{equation}
 \small
  \begin{aligned}
 {\mathcal F}_{\text{sup}}  =   \Big\{& a_0^{(1)} e^{j\theta_1} + a_0^{(2)} e^{j\theta_2} + \ldots  + a_0^{(d_f)}e^{j\theta_{d_f}} \\
 &  \quad \quad \quad \quad \vert \forall a_0^{(v)}\in \mathcal{A}_0, v = 1,2,\ldots,d_f \Big\}.
      \end{aligned}
 \end{equation}

 The superimposed constellation at each RE, i.e., $ {\mathcal F}_{\text{sup}} \in \mathbb C^{ M^{d_f} \times 1}$, has a significant impact on the error rate performance.  In addition, it is noted that the $d_f$ constellations at each RE are the same, i.e., $\mathcal{F}=\left \{
   \mathcal{A}_0{{e}^{j{{\theta }_{1}}}}, \mathcal{A}_0{{e}^{j{{\theta }_{2}}}}, \ldots, \mathcal{A}_0{{e}^{j{{\theta }_{d_f}}}}  \right \}$, which is referred to as  the benchmark constellation group.  However, the resultant constellations group $\mathcal{F}$ may not achieve the best error performance in different fading channels as the superimposed constellation $ {\mathcal F}_{\text{sup}} $ relies heavily on the choice of  $\mathcal{A}_0$  and the rotation angles. 
 Hence,  a jointly optimization scheme is proposed to generate the constellation group $\mathcal{F}$. Specifically, different from the above discussed scheme, where $\mathcal{F}$ is obtained by rotating a basic benchmark constellation $\mathcal{A}_0 $, we propose to directly design $\mathcal{F}= \left \{
   {{\mathcal{A}}_{1}},{{\mathcal{A}}_{2}},\cdots, {{\mathcal{A}}_{{{d}_{f}}}}\right \}$.   Define  the elements in ${{\mathcal{A}}_{v}}$ as ${{\mathcal{A}}_{v}} = \{a_{v,1},a_{v,2}, \ldots, a_{v,M} \}$.   According to the transmission model in the downlink fading channels, the constellation group $\mathcal{F}$ can be formulated as an optimization problem to minimize the SER of the superimposed constellation
\begin{equation}
\label{Opt1}
\begin{matrix}
  \left \{  
   {{\mathcal{A}}_{1}},{{\mathcal{A}}_{2}},\cdots, {{\mathcal{A}}_{{{d}_{f}}}}\right \}=\underset{0\le {{r}_{v,m}}\le 1,0\le {{\theta }_{v,m}}\le 2\pi }{\mathop{\arg \min }}\,{{P}_{b}}, \\ \\
    s.t.
    \begin{cases}
    {{P}_{b}} = {\frac{1}{{M}^{{{d}_{f}}}}}\sum\limits_{{{c}_{i}}\in {\mathcal F}_{\text{sup}}}{\sum\limits_{{{c}_{j}}\in {\mathcal F}_{\text{sup}}\backslash \left\{ {{c}_{i}} \right\}}{}}{{\left( 1+\frac{2\sigma _{Ra}^{2}{{E}_{s}}}{4{{N}_{0}}}{{\left| {{c}_{i}}-{{c}_{j}} \right|}^{2}} \right)}^{-1}},\\[6pt] 
    {{\mathcal{A}}_{v}} = \{a_{v,1},a_{v,2}, \ldots, a_{v,M} \}, \\[6pt] 
        {{a}_{v,m}}={{r}_{v,m}}{{e}^{j{{\theta }_{v,m}}}}, \\[6pt] 
      \mathcal F_{\text{sup}}= \left\{\sum\limits_{v=1}^{{{d}_{f}}}{{{a}_{v}}} \vert \;\forall {a}_{v} \in  {{\mathcal{A}}_{v}} \right\}, \\[6pt] 
    1\le i\le {{M}^{{{d}_{f}}}}, 1\leq v\leq d_f, 1\leq m\leq M&
    \end{cases}\\
\end{matrix}
\end{equation}

\noindent
where ${{P}_{b}}$ represents the SER of the superimposed codewords in the benchmark constellation group $\mathcal{F} $ under Rayleigh fading channel, and the SER can also be replaced for other fading channels.  ${c}_{i}$ and ${c}_{j}$ are two different superimposed codewords in ${\mathcal F}_{\text{sup}}$ where $i\neq j$, and $r_{v,m}$ and ${\theta_{v,m}}$ are the module and phase angle of the codewords $a_{v,m}$. Here, the  SQP  approach is adopted to solve the optimization problem. 

\begin{figure}[!t]
\centering
\subfloat[Chen's codebook \cite{On_the_Design_of_Near_Optimal_Sparse_Code_Multiple_Access_Codebooks} (${\mathcal F}^{\text{min}}_{\text{sup}}=0.28$).]{
\includegraphics[scale=0.7]{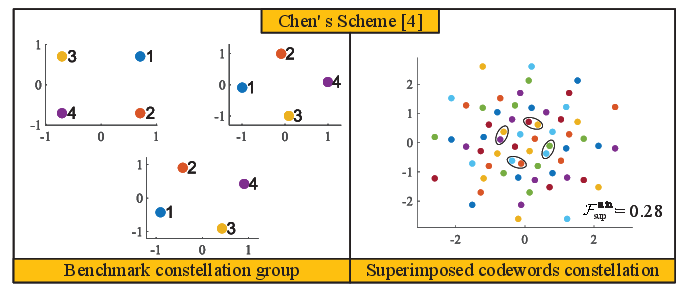}}\\
\subfloat[Proposed codebook (${\mathcal F}^{\text{min}}_{\text{sup}}=0.35$).]{
\includegraphics[scale=0.7]{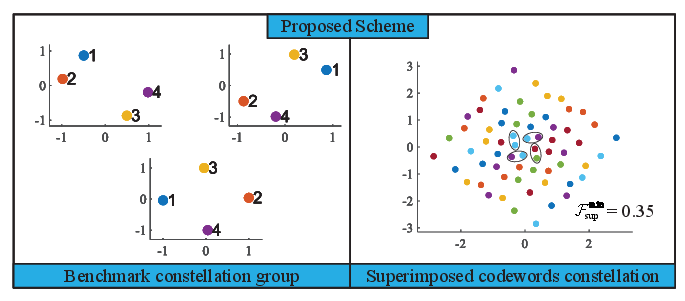}}
\caption{{Comparisons of the constellation group and superimposed codewords constellation of the proposed codebook and the codebook in \cite{On_the_Design_of_Near_Optimal_Sparse_Code_Multiple_Access_Codebooks}.}}
\label{fig:Comparisons of the constellation group}
\end{figure}

\textbf{\textit{Example 1:}}   Fig. \ref{fig:Comparisons of the constellation group} shows an example of the  benchmark constellation group $\mathcal{F} $ and the corresponding superimposed constellation $ {\mathcal F}_{\text{sup}}$ of the proposed scheme and   the existing scheme in\cite{On_the_Design_of_Near_Optimal_Sparse_Code_Multiple_Access_Codebooks}. In \cite{On_the_Design_of_Near_Optimal_Sparse_Code_Multiple_Access_Codebooks}, the  benchmark constellation group, as shown in Fig. \ref{fig:Comparisons of the constellation group}(a), is superimposed with  $d_f$   QPSK  constellations with certain rotations. In contrast, the benchmark constellation group generated by the proposed scheme is not confined to the rotation of a specific benchmark constellation $\mathcal{A}_{0}$. In addition, the proposed scheme achieves larger minimum Euclidean distance of the superimposed constellation, which is defined as 
{\setlength\abovedisplayskip{1pt}
\setlength\belowdisplayskip{1pt}
\begin{equation} 
\label{dmin_mc}
\small
{\mathcal F}^{\text{min}}_{\text{sup}}  \triangleq  \underset {{{c}_{i}}\in {\mathcal F}_{\text{sup}}, {{c}_{j}}\in {\mathcal F}_{\text{sup}}\backslash \left\{ {{c}_{i}} \right\}}{\min }    \vert  c_i-c_j\vert  ^2.
\end{equation}}

\subsection{Reconstruction of the Constellation Group}

\color{black}{ 
\begin{algorithm}[t] 
\caption{The Progressive Codebook Optimization Scheme.}
\label{alg:cap}
    \begin{algorithmic}[htbp]
    \STATE \textbf{Input:} REs $K$, users $L$, degree of user nodes $d_v$, degree of REs nodes $d_f$ , codebook size $M$.
    \STATE  \textbf{Return:} The factor graph matrix $\mathbf F$ through PEG.
    \STATE  \textbf{Optimize:}
    \STATE \hspace{\algorithmicindent}$\mathcal{F}\leftarrow$  Optimize (\ref{Opt1}) through SQP.
    \STATE \hspace{\algorithmicindent}${{\mathcal{F}}_{1}}\leftarrow$  Assign $\mathcal{F}$ randomly in the first RE.
    \FOR{$k=2$ to $K$}
        \STATE ${{\mathcal{F}}_{k}}\leftarrow$ Optimize {$\text { perms }(\cdot)$} and {${\mathbf{P}_{k,\phi_{k}^{v}}} $} in (\ref{Opt2}).
    \ENDFOR  
    \STATE  \textbf{Output:} The optimized multi-dimensional constellation.
\end{algorithmic}
\end{algorithm}}

\begin{figure}[!t]
	\centering
		\centering
		\includegraphics[width=1.0\linewidth]{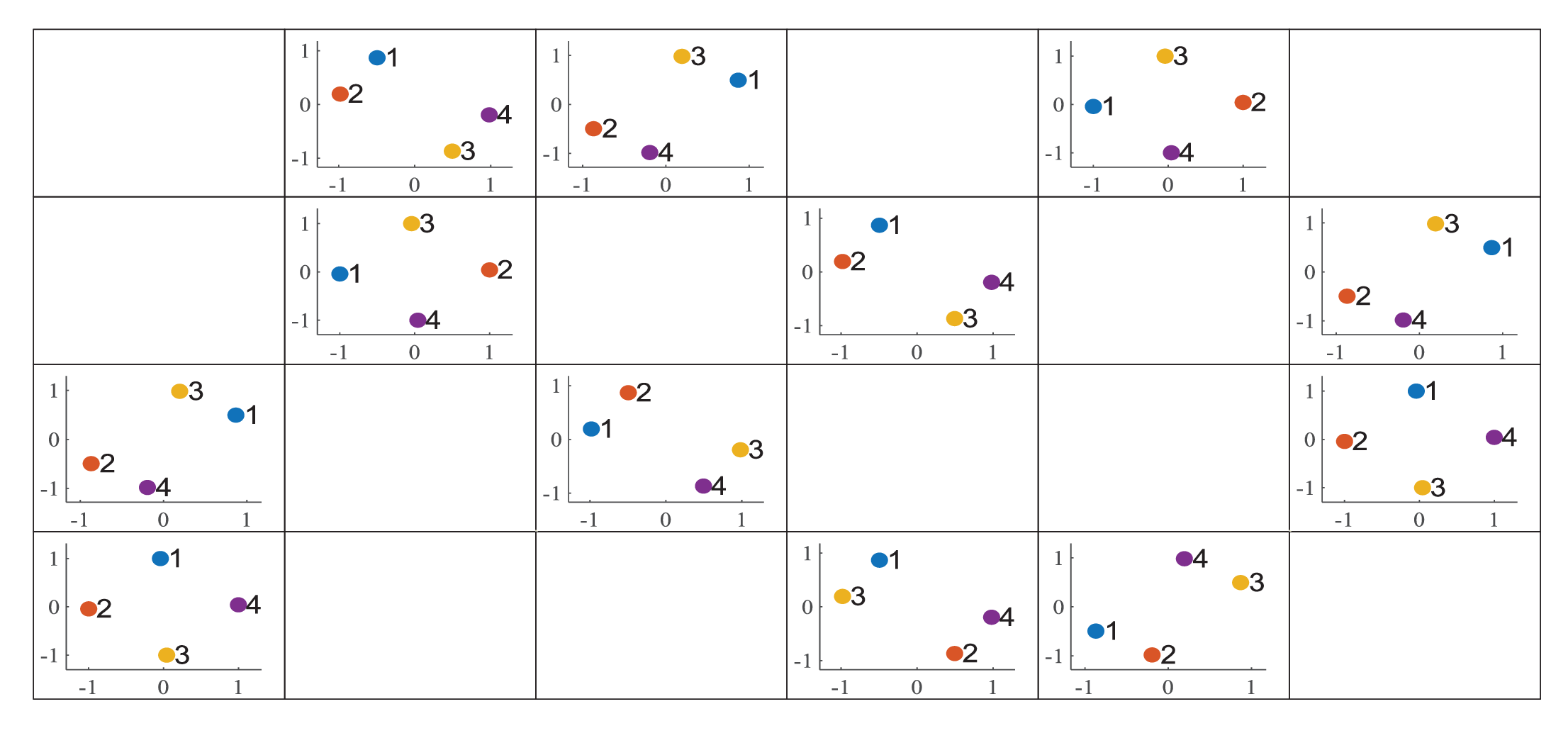}
        \caption{An example of the multi-dimensional constellation generated by  reconstructing the benchmark constellation group.}
        \label{fig:One Example of the Multidimensional Constellation Generated by the Proposed Codebook Design Scheme}
\end{figure}

After the benchmark constellation group  $\mathcal{F} $ is obtained, the multi-dimensional constellation for different users can be obtained by placing the sub-constellations of  $\mathcal{F}$ in the factor graph matrix to minimize the average SER of the superimposed codewords.
The non-zero constellation group at the first RE is trivial, which can be set the same as the optimized result $\mathcal{F}_{1}=\mathcal{F}$. However,  when $k\geq 2$, there are ${{d}_{f}}!$ possible combinational results of assigning the sub-constellations of   $\mathcal{F} $ to the activate users in each RE, where the activate users $f_{k,l}=1$ can be found through the indicator matrix $\textbf{F}$.  As $d_f$ generally is a small value, we adopt the ergodic search method  to obtain the best result. Besides, the labeling sequence of each sub-constellation ${{\mathcal{A}}_{i}}$ in $\mathcal{F}$ is reordered to maximize the coding gains of the codebook \cite{SCMA_Codebook_for_Uplink_Rician_Fading_Channels}, where the ergodic search method is employed to achieve the best result . Finally, the problem of reconstructing the benchmark constellation group $\mathcal{F}$ in the $k\text{th}$ RE can be formulated as an optimization problem to minimize the SER of the superimposed codewords in previous $k$ REs. Denote the reconstructed constellation group in the $k\text{th}$ RE as ${\mathcal F}_{k} = \left \{
   {{\mathcal{S}}_{k,\phi_{k}^{1}}},{{\mathcal{S}}_{k,\phi_{k}^{2}}},\cdots, {{\mathcal{S}}_{k,\phi_{k}^{d_f}}}\right \}$, where ${{\mathcal{S}}_{k,\phi_{k}^{v}}}$ denotes the reconstructed constellation for the $ \phi_{k}^{v}$th  user at the $k$th RE, and $ \phi_{k}^{v}$ denotes the   active user index in the $k$th RE. For example, for the first RE, i.e. $k=1$, we have   $\phi_{1}  =[2,3,5]$ and $\phi_{1}^{1} = 2$. Namely,   ${{\mathcal{S}}_{1,\phi_{1}^{1}}}$ is assigned for the second user at the first RE.
 In Rayleigh fading channels,  the design of reconstructed constellation group ${\mathcal F}_{k}$ is formulated as 
\begin{equation}
\label{Opt2}
\begin{aligned}
\begin{array}{l} {\mathcal F}_{k} 
     =  \underset{\text { perms }(\cdot), \mathbf{P}_{k,l}}{\arg \min } \\
     \sum\limits_{c_{k,i} \in {\mathcal F}_{\text{sup}}^k} \sum\limits_{\substack{c_{k,j} \in {\mathcal F}_{\text{sup}}^k \\\backslash\left\{c_{k, i}\right\}}} \prod\limits_{n=1}^{k}\left(1+\frac{2 \sigma_{R a}^{2} E_{s}}{4 N_{0}}\left|c_{n, i}-c_{n, j}\right|^{2}\right)^{-1}, \\ \\
    s.t. =
    \left\{ \begin{array}{l}
    \left \{ 
    {{\mathcal{S}}_{k,\phi_{k}^{1}}},{{\mathcal{S}}_{k,\phi_{k}^{2}}}, \cdots, {{\mathcal{S}}_{k,\phi_{k}^{d_f}}}
     \right  \} =  {\rm{perms}} \\
    \left( {\left \{  
    {{{\cal A}_1}{{\bf{P}}_{k,1}}}, {{{\cal A}_2}{{\bf{P}}_{k,2}}},\ldots, {{{\cal A}_{{d_f}}}{{\bf{P}}_{k,d_f}}} \right \}} \right), \\  \\
  \mathcal F_{\text{sup}}^k= \left\{\sum\limits_{v=1}^{{d_f}} {{{s}_{k,v}}} \vert \;\forall {s}_{k,v} \in    {{\mathcal{S}}_{k,\phi_{k}^{v}}} \right\}, \\\\ 
    2 \leq k \leq K, 
    1 \leq v \leq d_f,
 \end{array}\right. 
\end{array}
\end{aligned}
\end{equation}
 
\noindent
where ${\mathcal F}_{\text{sup}}^k $ denotes the superimposed constellation  at the $k\text{th}$ RE, $ {{\mathcal{S}}_{k,\phi_{k}^{v}}}\in \mathbb C^{M \times 1}$ represents the constellation of the  $\phi_{k}^{v}\text{th}$ user in the $k\text{th}$ RE, and the constellations of the inactivate users $f_{k,l}=0$ who does not perform transmission in the $k$th RE is set to $\textbf{0}$ with $M$ zero elements,  $\text{perms}\left( \cdot \right)$ denotes the permutation function  for generating all possible combinations of the  constellation group,  and  the binary matrix  ${\mathbf{P}_{k,v}} $ denotes permutation matrix that reorders the labelling sequence of a constellation. Specifically,  ${\mathbf{P}_{k,v}} $ is defined as  ${\mathbf{P}_{k,v}}= \left\{ p_{m,n} \right\}^{M \times M} $, where $p_{m,n} \in \{0,1\}$ and $p_{m,n}=1$ denotes the $m$th codeword is labeled with decimal $n$. There are    $M!$ permutation (labeling) options for  an $M$-ary  constellation. Hence, for large modulation order $M$, the complexity of  exhaustive search is prohibitively high.   The binary switching algorithm employed in \cite{LuoLPSCMA} can be utlized   to find the labeling solution with a reasonable complexity.  \color{black}   Finally, the detail design of the  proposed codebook is summarized in   \textbf{Algorithm \ref{alg:cap}}.

\textbf{\textit{Example 2:}} We now present an example to illustrate the proposed constellation reconstruction process. Specifically, consider the  benchmark constellation group  presented in Fig. \ref{fig:Comparisons of the constellation group}, the constructed codebook for different users is given as follows  
\begin{equation}
\small
\label{exCB}
\setlength{\arraycolsep}{1pt}
\begin{aligned}
   \begin{matrix}
  {\mathcal X}\\
    \end{matrix}  & =
    \left[ \begin{matrix}
   \textbf{0} & {{\mathcal{S}}_{1,2}} & {{\mathcal{S}}_{1,3}} & \textbf{0} & {{\mathcal{S}}_{1,5}} & \textbf{0}  \\
   \textbf{0} & {{\mathcal{S}}_{2,2}} & \textbf{0} & {{\mathcal{S}}_{2,4}} & \textbf{0} & {{\mathcal{S}}_{2,6}}  \\
   {{\mathcal{S}}_{3,1}} & \textbf{0} & {{\mathcal{S}}_{3,3}} & \textbf{0} & \textbf{0} & {{\mathcal{S}}_{3,6}}  \\
   {{\mathcal{S}}_{4,1}} & \textbf{0} & \textbf{0} & {{\mathcal{S}}_{4,4}} & {{\mathcal{S}}_{4,5}} & \textbf{0}  \\
\end{matrix} \right],\\
 & =  \left[ \begin{matrix}
   \textbf{0} &  {{{\cal A}_1}{{\bf{P}}_{1,1}}} &  {{{\cal A}_2}{{\bf{P}}_{1,2}}} & \textbf{0} & {{{\cal A}_3}{{\bf{P}}_{1,3}}} & \textbf{0}  \\
   \textbf{0} & {{{\cal A}_3}{{\bf{P}}_{2,3}}} & \textbf{0} & {{{\cal A}_1}{{\bf{P}}_{2,1}}}  & \textbf{0} & {{{\cal A}_2}{{\bf{P}}_{2,2}}}  \\
   {{{\cal A}_2}{{\bf{P}}_{3,2}}}  & \textbf{0} & {{{\cal A}_1}{{\bf{P}}_{3,1}}}  & \textbf{0} & \textbf{0} & {{{\cal A}_3}{{\bf{P}}_{3,3}}}   \\
 {{{\cal A}_3}{{\bf{P}}_{4,3}}}  & \textbf{0} & \textbf{0} & {{{\cal A}_1}{{\bf{P}}_{4,1}}}  & {{{\cal A}_2}{{\bf{P}}_{4,2}}}  & \textbf{0}  \\
\end{matrix} \right],
\end{aligned}
\end{equation}
where the corresponding constellations are also illustrated in Fig. \ref{fig:One Example of the Multidimensional Constellation Generated by the Proposed Codebook Design Scheme}. The row represents the RE and the column denotes the user. In (\ref{exCB}), the permuted  ${\cal A}_2$ and ${\cal A}_3$ are assigned for the first user, and the codebook for the remaining  users can also be generated similarly based on  (\ref{exCB}). 
In addition, it can be observed from Fig. \ref{fig:One Example of the Multidimensional Constellation Generated by the Proposed Codebook Design Scheme} that the multi-dimensional constellation of each user is not restricted to the operation of a specific benchmark constellation. For example, the constellation of users $2$ and $ 3$ on the first RE is different from user $5$. In addition, the permutation function of the constellation and the relabeling sequence of the constellation can be noticed to increase the coding gains of the codebook.

\color{black} 

\section{Simulation Results}

\begin{figure*}[htbp]
	\centering
	\begin{minipage}{0.49\linewidth}
		\centering
		\includegraphics[width=1.0\linewidth]{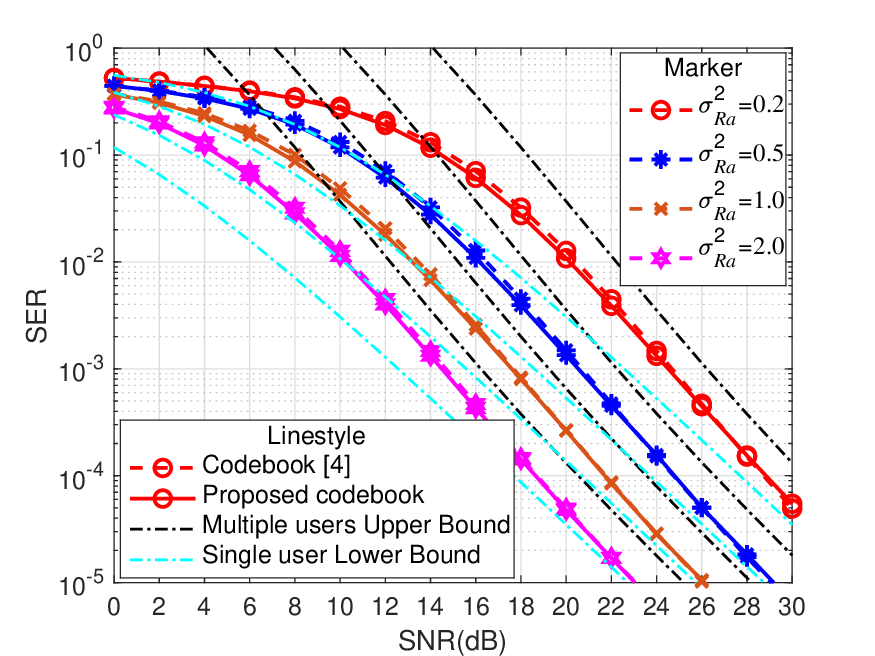}
        \caption{The performance comparisons between the proposed codebooks and Chen's codebook \cite{On_the_Design_of_Near_Optimal_Sparse_Code_Multiple_Access_Codebooks} in Rayleigh fading channels.}
        \label{fig:simulation_fig1}
	\end{minipage}
	\begin{minipage}{0.49\linewidth}
		\centering
		\includegraphics[width=1.0\linewidth]{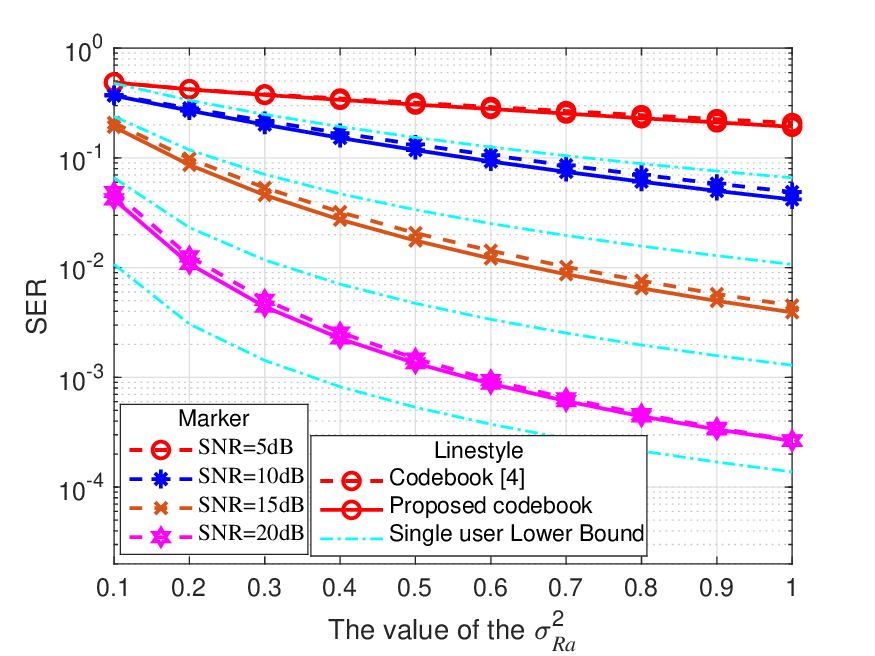}
        \caption{The SER performance of the proposed codebooks and Chen's  codebook \cite{On_the_Design_of_Near_Optimal_Sparse_Code_Multiple_Access_Codebooks} in different Rayleigh fading parameters.}
        \label{fig:simulation_fig2}
	\end{minipage}
\end{figure*}

In this section, we  present the SER performance of the proposed SCMA codebooks in comparison with the one introduced in \cite{On_the_Design_of_Near_Optimal_Sparse_Code_Multiple_Access_Codebooks}, which has been verified to yield good error performance in downlink Rayleigh channels \cite{A_Novel_Scheme_for_the_Construction_of_the_SCMA_Codebook}, \cite{A_Comprehensive_Technique_to_Design_SCMA_Codebooks}. Then, simulation results are provided to demonstrate that the codebooks designed by progressively minimizing the SER of product distance of the superimposed codewords outperform the benchmark  codebook \cite{On_the_Design_of_Near_Optimal_Sparse_Code_Multiple_Access_Codebooks} under different downlink fading channels at lower SNRs.
\subsection{Error Perforamnce in Rayleigh Channel}

 Fig. \ref{fig:simulation_fig1} illustrates the SER performance of the codebooks in Rayleigh fading channels with different scatter components $\sigma _{Ra}^{2}$, where the single user bounds are obtained by calculating the average SER of the mother constellation of each user.  It can be observed that the performance of the SCMA codebooks approaches the single-user error performance at higher  SNRs, indicating a gradual mitigation  of interference among users.  Since maximizing the product distance of the superimposed codewords ${{d}_{p}}$ is  crucial  only when the values of SNR and  $\sigma _{Ra}^{2}$ are large, we can observe from Fig. \ref{fig:simulation_fig1} that the proposed codebooks outperform Chen's codebook \cite{On_the_Design_of_Near_Optimal_Sparse_Code_Multiple_Access_Codebooks} at low SNRs. Specifically, the proposed codebook achieves about    $0.3$ dB gains over Chen's codebook at SER = $10^{-2}$ when the scatter component $\sigma _{Ra}^{2}=0.2$.

  Fig. \ref{fig:simulation_fig2} presents  the  SER of the codebooks with different   values of the scatter component $\sigma _{Ra}^{2}$   under various SNRs. As evidenced by  Fig. \ref{fig:simulation_fig2}, better SER performance can be achieved with the increase of the scatter component $\sigma _{Ra}^{2}$. In addition, Fig. \ref{fig:simulation_fig2} also demonstrates that the proposed codebooks obtained by minimizing the SER of the SCMA achieve  better error performance than Chen's codebook   under lower SNRs and scatter components $\sigma _{Ra}^{2}$.

\subsection{Error Performance in Rician Channel}

\begin{figure*}[htbp]
	\centering
	\begin{minipage}{0.49\linewidth}
		\centering
		\includegraphics[width=1.0\linewidth]{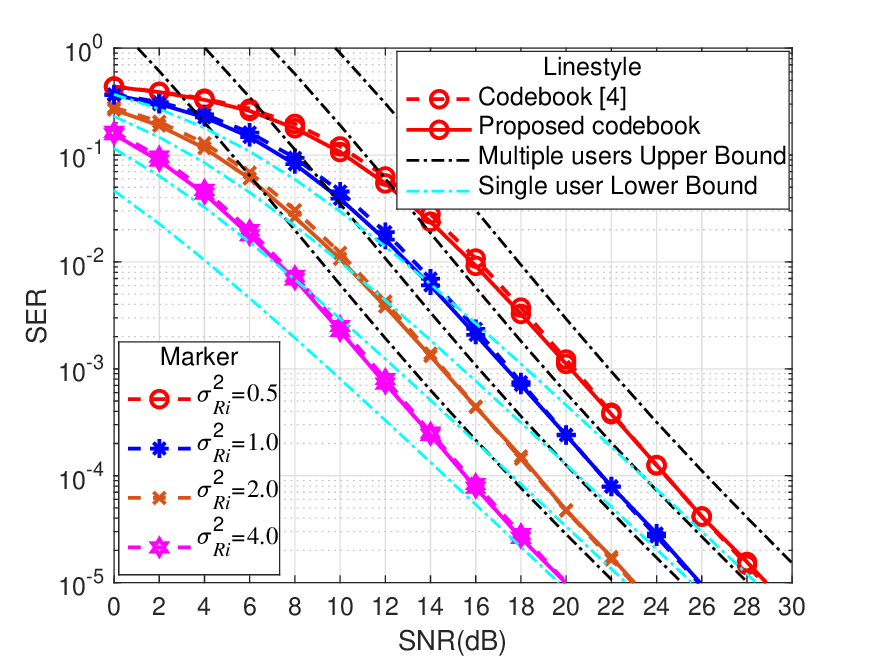}
		\caption{The SER performance of the proposed codebooks and Chen's codebook \cite{On_the_Design_of_Near_Optimal_Sparse_Code_Multiple_Access_Codebooks} in the Rician fading channel under various scatter components $\sigma _{Ri}^{2}$ where the LoS component $u=0.2$.}
        \label{fig:simulation_fig3}
	\end{minipage}
	\begin{minipage}{0.49\linewidth}
		\centering
		\includegraphics[width=1.0\linewidth]{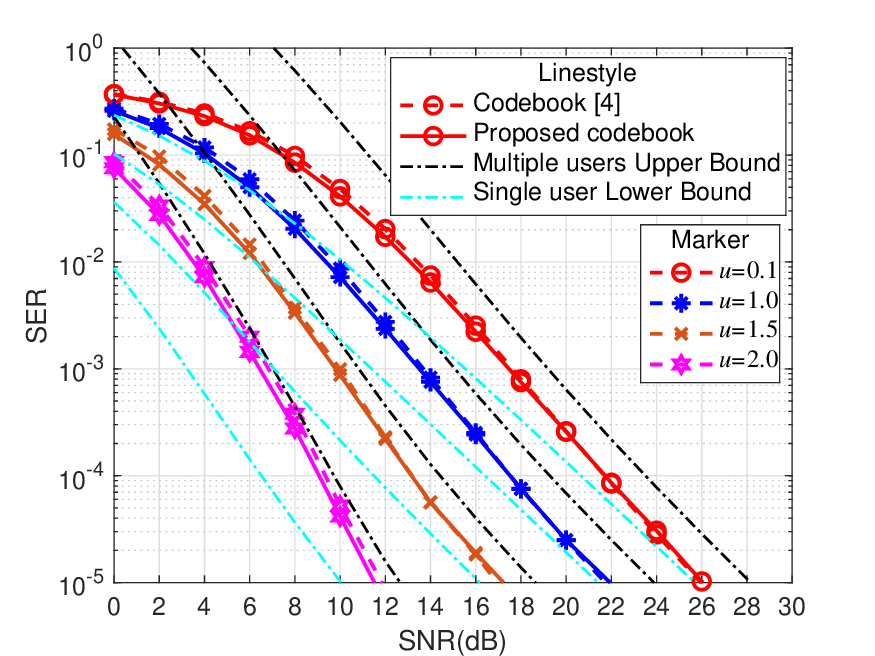}
		\caption{The SER performance of the proposed codebooks and Chen's  codebook \cite{On_the_Design_of_Near_Optimal_Sparse_Code_Multiple_Access_Codebooks} in the Rician fading channel under various LoS components $u$ where the scatter component $\sigma _{Ri}^{2}=0.5$.}
    \label{fig:simulation_fig4}
	\end{minipage}
\end{figure*}

 Fig. \ref{fig:simulation_fig3} compares the SER performance of the proposed codebook and Chen's codebook  under Rician fading channels with various scatter component  $\sigma _{Ri}^{2}$. At low SNRs, the Euclidean distance of the superimposed codewords ${{d}_{e}}$ is  one of important factors affecting the error performance. Consequently,  a gain of around $0.2$ dB can be observed when compared with Chen's codebook. However, in the case of  higher SNRs, the proposed codebooks achieve the same SER performance as Chen's codebook. This is because   the key factor affecting the SER performance at  higher SNRs is maximized  product distance of the superimposed codewords ${{d}_{p}}$.
 
 Fig. \ref{fig:simulation_fig4} presents the SER of the proposed codebooks under different LoS components $u$. In the case of a   small value of $u$, i.e., $u=0.1$, the proposed codebook achieves the same SER performance as Chen's codebook \cite{On_the_Design_of_Near_Optimal_Sparse_Code_Multiple_Access_Codebooks} at a higher SNR,  while a slightly better error performance can be observed at lower SNRs. As $u$ increases, i.e., when the LoS dominates,  the proposed codebooks achieve  better SER performance than Chen's codebook \cite{On_the_Design_of_Near_Optimal_Sparse_Code_Multiple_Access_Codebooks}. Specifically, about $ 0.3$ dB gain can be observed  at SER = $10^{-4}$ for the LoS component $u=2.0$. This is because the design criteria of maximizing the Euclidean distance of the superimposed codewords ${{d}_{e}}$ is also a crucial factor influencing the error performance of SCMA under a larger  LoS  component $u$. Moreover, as the LoS component $u$ increases, the Rician factor becomes larger, resulting in better error performance that can be achieved from the LoS component  ${{\Pr }_{\text{LoS}}}$.

\subsection{Error Performance in Nakagami-$m$ Channel}

\begin{figure*}[htbp]
	\centering
	\begin{minipage}{0.49\linewidth}
		\centering
		\includegraphics[width=1.0\linewidth]{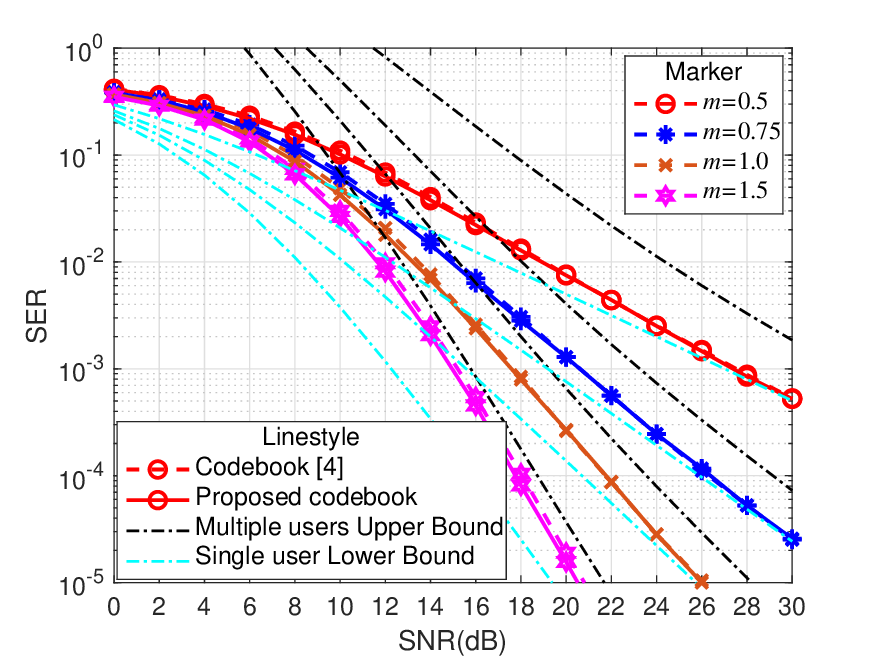}
		\caption{The SER performance of the proposed codebooks and Chen's  codebook \cite{On_the_Design_of_Near_Optimal_Sparse_Code_Multiple_Access_Codebooks} in Nakagami-$m$ fading channel under various Nakagami-$m$ fading parameters $m$ where the scale parameter is $\text{ }\!\!\Omega\!\!\text{ }=1$.}
        \label{fig:simulation_fig5}
	\end{minipage}
	\begin{minipage}{0.49\linewidth}
		\centering
		\includegraphics[width=1.0\linewidth]{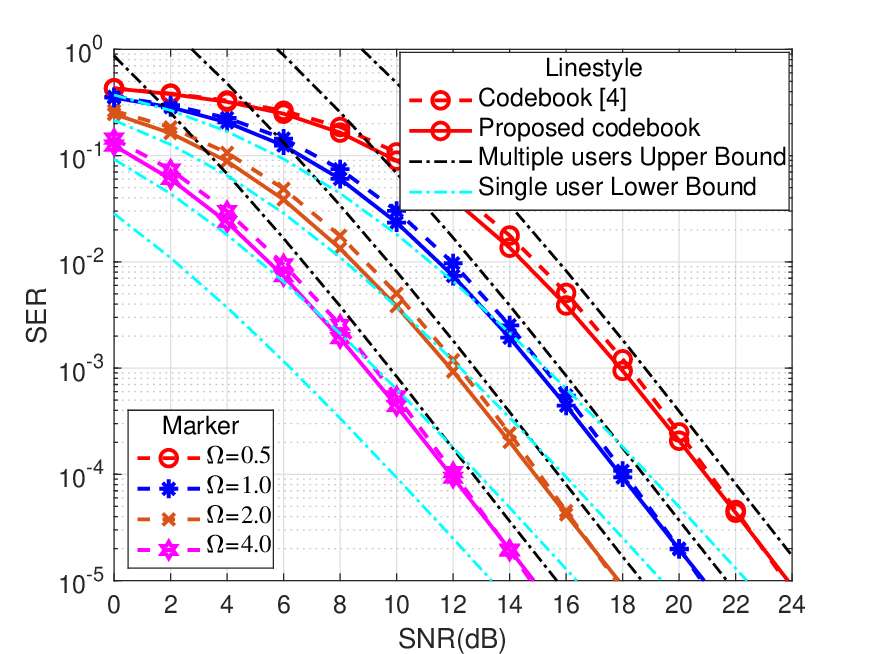}
		\caption{The SER performance of the proposed codebooks and Chen's codebook \cite{On_the_Design_of_Near_Optimal_Sparse_Code_Multiple_Access_Codebooks} in Nakagami-$m$ fading channel under various scale parameters $\text{ }\!\!\Omega\!\!\text{ }$ where the Nakagami-$m$ fading parameters is $m=1.5$.}
    \label{fig:simulation_fig6}
	\end{minipage}
\end{figure*}

Fig. \ref{fig:simulation_fig5} and Fig. \ref{fig:simulation_fig6} compare the SER performance of the proposed codebook and Chen's codebook under different Nakagami-$m$ fading channels. As can been seen from Fig. \ref{fig:simulation_fig5}, the proposed codebook outperforms Chen's codebook \cite{On_the_Design_of_Near_Optimal_Sparse_Code_Multiple_Access_Codebooks} at low SNRs. For the Nakagami-$m$ fading channels with   $m=1$, which corresponds to  the Rayleigh fading channel, the proposed codebooks achieve  the same SER performance as Chen's codebook \cite{On_the_Design_of_Near_Optimal_Sparse_Code_Multiple_Access_Codebooks} at higher SNRs. As $m$ increases, the ratio of the ${\text{ }\!\!\Omega\!\!\text{ }}/{m}\;$ decreases. Therefore, maximizing the Euclidean distance of the superimposed codeword ${{d}_{e}}$ can help improve the SER performance. When the  Nakagami-$m$ fading parameters are $m=1$ and $m=1.5$, the proposed codebooks achieve  approximately 0.1 dB and 0.2 dB over Chen's codebook at SER = $10^{-3}$.

  Fig. \ref{fig:simulation_fig6} compares the SER   performance of different codebooks   under various scale parameters $\text{ }\!\!\Omega\!\!\text{ }$. It can be seen that as $\text{ }\!\!\Omega\!\!\text{ }$ increases , the error performance of the SCMA improves. Moreover, the increased ratio of ${\text{ }\!\!\Omega\!\!\text{ }}/{m}\;$ indicates that the key performance indicator lies in maximizing the product distance of the superimposed codewords ${{d}_{p}}$.    Thus, the proposed codebooks exhibit a comparable  SER performance to Chen's codebook  with the increase of the scale parameters $\text{ }\!\!\Omega\!\!\text{ }$ at high SNRs. However, in lower SNRs, around $0.3$ dB   gain  is observed for the proposed codebook    at SER = $10^{-3}$.

\section{Conclusion}
{
In this paper, we have proposed a progressive codebook optimization scheme for downlink  SCMA  systems over fading channels. Utilizing the  MGF, we have derived a simplified  SER  model of SCMA under various channel conditions. This model serves as the design criterion for optimizing the codebook. Subsequently, we have optimized the benchmark constellation group at a single RE  using a  SQP  approach. Additionally, we have introduced a constellation group reconstruction to progressively assign the benchmark constellation group in each RE. For the current RE, the assignment of sub-constellations was  designed by minimizing the error performance of the product distance of superimposed codewords from previous REs. Finally, extensive simulation results have been conducted to demonstrate the superiority of the proposed codebooks at low SNRs in comparison to the benchmark codebooks over different fading channels.}

\bibliography{ref} % put your favourite Bibtex archive references here
\bibliographystyle{IEEEtran}


%\end{IEEEbiography}


%\newpage



%\vfill


\end{document}